\documentclass[a4paper]{VisionStyle}
\usepackage{epsfig}

\begin{document}

\title{Magnetic flares and outflows from structured accretion disc coronae}

\author{A.\,Merloni\inst{1}\inst{2} \and A.C.\,Fabian\inst{1} } 

\institute{
Institute of Astronomy, Madingley Road, Cambridge, CB3 0HA, United Kingdom
\and 
  Max-Planck Institute f\"ur Astrophysik, Karl-Schwarzschild Str. 1,
Garching, D-85741, Germany}

\maketitle 

\begin{abstract}
We present a model for magnetic structured coronae above accretion
discs. 
On the shortest timescales, spatially and temporally correlated
coronal flares can explain X-ray temporal and spectral variability
observed in Seyfert galaxies. 
In particular, power density spectra, flux-spectral index and flux-variance correlations
are naturally accounted for by the model. 
More dramatic spectral variations (i.e. state transitions in GBHC) are 
associated with parameters varying on longer timescales, such as accretion rate, 
coronal strength  or geometry of the inner disc. In the framework of
the standard Shakura--Sunyaev accretion disc theory, here we discuss why energetically 
dominant coronae at low accretion rates are ideal sites for launching powerful 
MHD driven outflows. Then, if the outflow is radiatively 
inefficient, then so is the source overall, even without advection
being relevant  for the dynamics of the accretion flow. This could be an alternative 
scenario for LLAGN and GBHC in their low/hard state, and may have
consequences for our understanding of the accretion history of the universe.
\keywords{accretion, accretion discs - black hole physics - magnetic fields}
\end{abstract}

\section{Introduction}

The hard X-ray spectra, the properties of the reflection features and
of rapid time variability in accreting black holes can all be
considered as indications
of the presence of a structured, hot, optically thin component in the
inner part of the accretion flow (the {\it corona}), situated above a cold, geometrically
thin, disc. Many models of the accretion disc corona have been
proposed in recent years, which are able to fit the observed {\it time
averaged} spectra in terms of Comptonization of soft photons in the
hot corona. However, many uncertainties regarding the actual geometry of
the inner accretion flow remain unsolved (see Done 2001 for a
review). Here we
present a model for a structured, magnetic corona above a standard,
geometrically thin and optically thick disc, that can satisfy
observational constraints on both long and short time variability
of these systems.

Let us briefly sketch the overall energetics of an accretion disc-corona
system.
The total power released by the accreting gas is
defined as $L\equiv \dot m L_{\rm Edd}=4\pi G M m_{\rm p}\dot m
c/\sigma_{\rm T}$. To be fairly general, we assume that the coronal
power generated by the disc (which is a fraction $f$ of the total: $L_{\rm c}=f\dot m L_{\rm Edd}$)
can be either dissipated locally to heat the corona, and ultimately radiated away
as hard X-rays with a luminosity $L_{\rm H}=(1-\eta)L_{\rm c}$, or
 used to launch an outflow with power $L_{\rm j}=\eta L_{\rm c}$. 
The main characteristic of our  model are the following:
in the hot corona the energy is stored in a strong,
highly intermittent magnetic field (amplified in the turbulent disc
and buoyantly expelled in the vertical direction).
Magnetic energy dissipation occurs at the smallest end of the
turbulent-energy cascade. Such small flares heat the corona (with a
power $L_{\rm H}$) and can trigger an avalanche
in their immediate neighborhood (\cite{amerloni-C2:pf99}), creating a bigger active region, and
producing the observed flares in the lightcurves. In Sec.~\ref{amerloni-C2_sec_short}, a
stochastic model (the so-called thundercloud model, Merloni \& Fabian,
2001, MF1) of the short-time variability is discussed.

On longer timescales, variations of the fraction of power released into the corona, $f$, and of the
fraction of coronal power used to launch an outflow, $\eta$, 
may be associated, for example, to changes of the accretion
rate and/or of the geometry of the inner disc. 
In Sec.~\ref{amerloni-C2_sec_long} we show that, under reasonable assumptions on the
nature of the disc viscosity, the strength of the corona $f$ increases as
the accretion rate decreases. Also we discuss reasons why 
an energetically dominant corona can be the
site where powerful outflows are produced. 

\section{Short timescales: magnetic flares}
\label{amerloni-C2_sec_short}
Seyfert 1 galaxies show strong temporal variability in the X-ray band, 
with a Power Density Spectrum (PDS) characterized by a ``red noise''
power-law: 
$P(f)=Af^{-\alpha}$ with $\alpha \sim 1.5$ (see e.g. Nandra, 2001).
The dynamical timescale of such systems, 
$t_{\rm dyn}=9\times 10^3 r^{3/2} (M_{\rm
BH}/10^8 M_{\sun})$~s, is such that good time-resolved 
spectroscopy can directly provide information about the 
heating mechanism of the corona. 
In fact, from the observational point of view 
there is a growing evidence that during a flare the X-ray
spectrum becomes softer as the (2-10 keV) flux increases (see
e.g. Vaughan \& Edelson 2001).
Also, a remarkable linear correlation between variability and flux on
all timescales has recently been reported (Uttley \& Mc Hardy 2001).

The coronal magnetic field intensity depends on the dissipation rate and on the
field geometry. Here we assume, on the basis of the strong spectral
and temporal variability,
that  the corona is patchy and the magnetic field concentrated in a number of
small active regions where dissipation occurs  (\cite{amerloni-C2:hmg94,amerloni-C2:ste95,amerloni-C2:dcf99}).

\begin{figure}[ht]
\vspace{-0.6cm}
  \begin{center}
    \epsfig{file=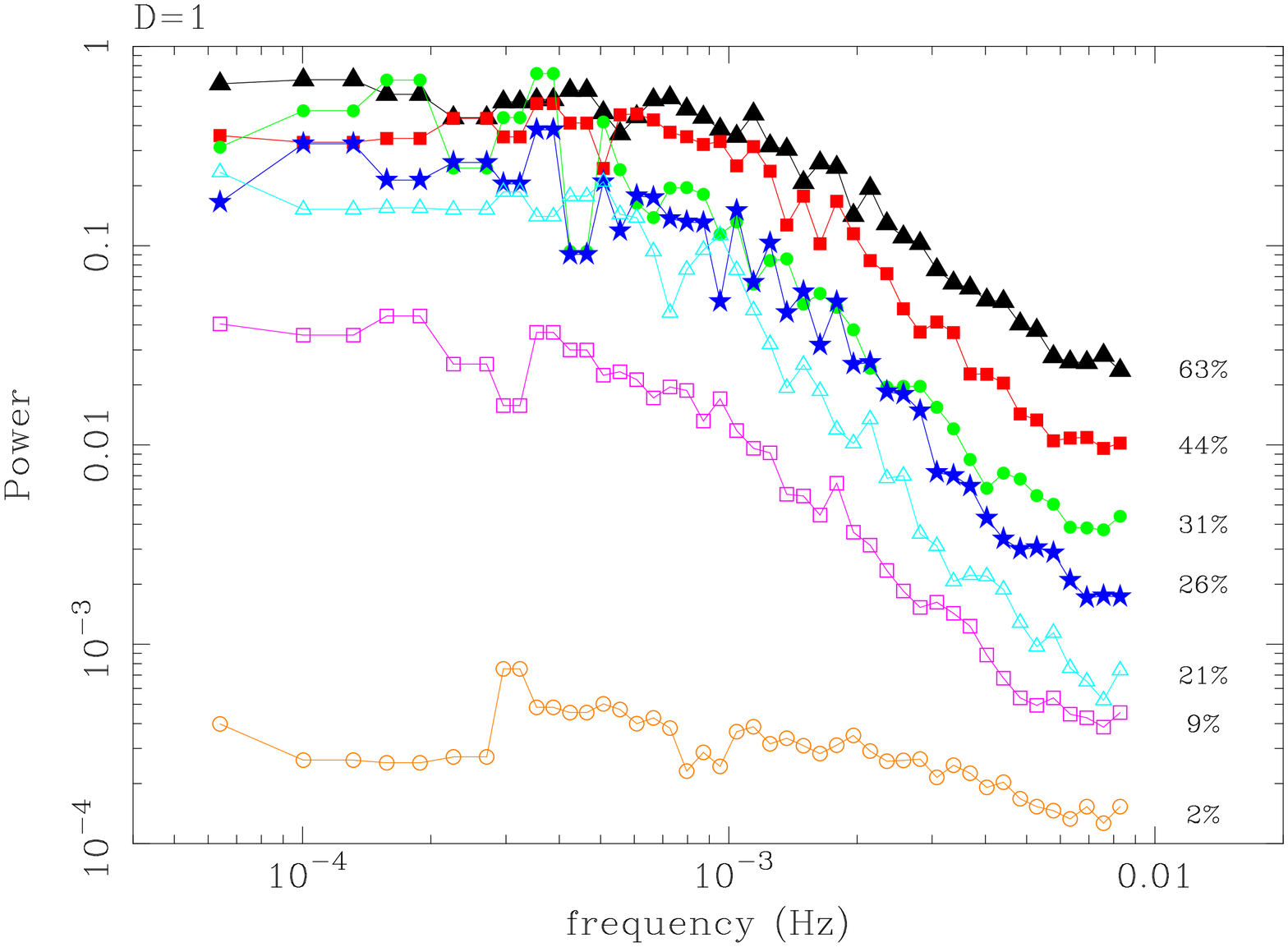, width=8cm}
  \end{center}
\caption{Average PDS of ten segments of a simulated lightcurve (each 
$3\times 10^4$~seconds long), for fixed coronal optical depth,
$\tau=1.5$ and $D=1$. From bottom to top the different curves
correspond to a covering fraction $C=0.03$ (orange, open circles), $0.01$ 
(purple, open squares), $0.005$ (cyan, open triangles), $0.002$ 
(blue, filled stars), $0.001$ (green, filled circles), $0.0003$ 
(red, filled squares) and $0.0001$ (black, filled triangles). 
On the right, close to each curve, the corresponding 
fractional rms is reported.}
\label{amerloni-C2_pds}
\end{figure}

We assume that the luminosity of an active region  is determined by
the size of the avalanche, and therefore scales as 
the size of the region: $L(R_a)=C_1 R_a^D$. 
Due to the fractal nature of  reconnection and avalanche generation, 
we also assume that the number of active regions of size $R_a$ 
at any time, is given by $dN(R_a)/dR_a=C_2 R_a^{-p}$.
The PDS slope, $\alpha$, is then related to the indices $D$ and $p$ by
the relation $p= 2D + 3 -\alpha$.
The constants $C_1$ and $C_2$ can be fixed imposing the overall
normalization for the total corona covering fraction and the average
hard luminosity (see MF1 for details).	
The covering fraction $C$ is related to the total variability level as
illustrated in the example in Fig.~\ref{amerloni-C2_pds}. 

As discussed in MF1, we have simulated AGN X-ray lightcurves fixing $\alpha = 1.5$ (in order to
reproduced the observed properties of the PDS), 
and studied the time and spectral variability properties of the model.  		

\begin{figure*}
\vspace{-0.7cm}
\begin{center}
    \epsfig{file=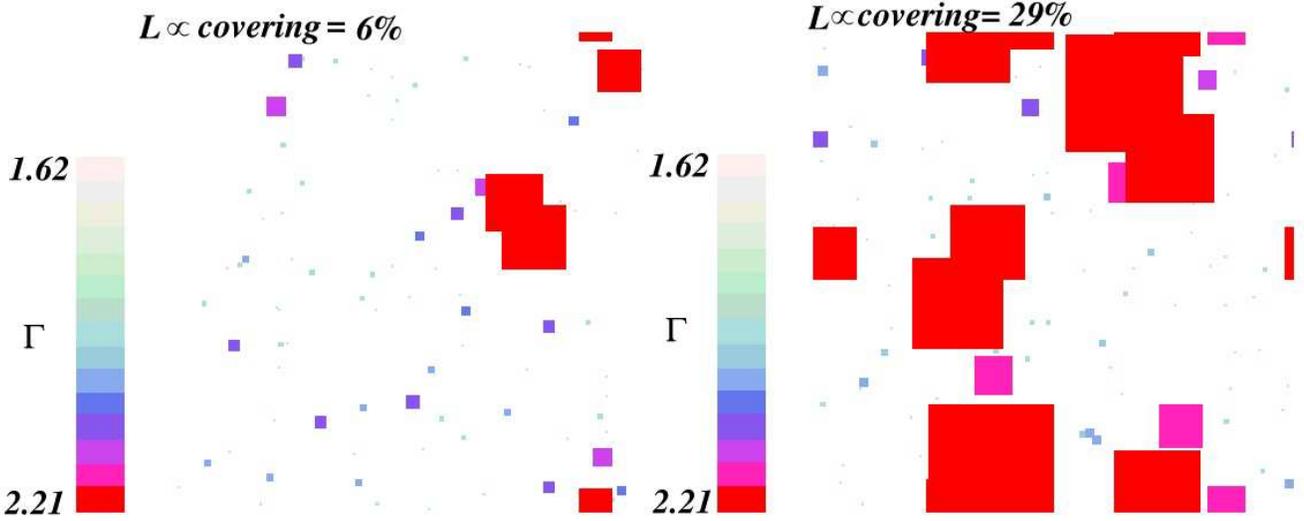}
  \end{center}
\vspace{-2.1cm}
\caption{The inner corona is mapped into a square. For simplicity, we
chose $D=2$, so that the pixels have equal luminosity, and the total
luminosity scales with the covered area. The region of the corona
active at any time are represented by filled squares, whose size is
distributed as a power-law, with index $p=2$. Each region is assumed
to have the same height above the disc ($H_a=R_{\rm S}$) and optical
depth ($\tau=1$). The color code shows the spectral index of the
filled active regions, calculated self-consistently taking into
account thermal Comptonization of disc photons (both intrinsic and
reprocessed) and of synchrotron emission in the active regions
themselves (see MF1). It ranges from $\Gamma=1.62$ (pink) to $\Gamma=2.21$ (red). Clearly the more luminous state (right hand side, for which the instantaneous covering fraction is larger), is dominated by active regions with soft spectra.
}
\label{amerloni-C2_map}
\end{figure*}

We assume that the X-ray spectrum is produced by thermal
Comptonization 
in each spherical active region of size $R_a$, 
lifted above the disc at a height $H_a$ and with Thomson optical depth
$\tau$. 
The active regions illuminate
the cold disc: 
part of the flux is reflected and part is absorbed and reprocessed. 
We assume that most of the accretion power is dissipated in the corona
($f > 0.7$), so that reprocessed radiation is the main source 
of soft photons in the active regions. 
For large and luminous active regions 
the amount of soft radiation intercepted by an active region
increases, and the spectrum is softer. This is illustrated in the
example of Fig.~\ref{amerloni-C2_map}.

We calculate the spectrum using an analytic approximation for 
thermal Comptonization (see MF2), fixing $H_a=R_{\rm S} =2GM_{\rm
BH}/c_2$, 
and we study the correlations between spectral properties 
and total luminosity (in the 3-10~keV band). 
We have fixed $M_{\rm BH}=10^7M_{\sun}$ and an accretion rate of 10\% of the 
Eddington one. In MF1 we show in detail how such a model can reproduce
the observed luminosity - spectral index correlation. In the following
section, instead, we will focus on the luminosity-variability
correlation (see also Merloni 2001).
 
\subsection{The luminosity-variability correlation}
Recently, \cite*{amerloni-C2:umh01} have discovered a remarkable linear
correlation, 
on all timescales (Uttley, private communication), between
 variability and flux in many compact accreting systems.

We have analysed one of our simulated lightcurves in order to look for a similar correlation.
  With the parameters $D$ and $C$ fixed to the values of $1$ and $0.001$, respectively,
we have produced a lightcurve of the total duration  of about $3 \times 10^6$ s, 
with resolution of 1 minute. 

\begin{figure}[ht]
\vspace{-0.6cm}
  \begin{center}
    \epsfig{file=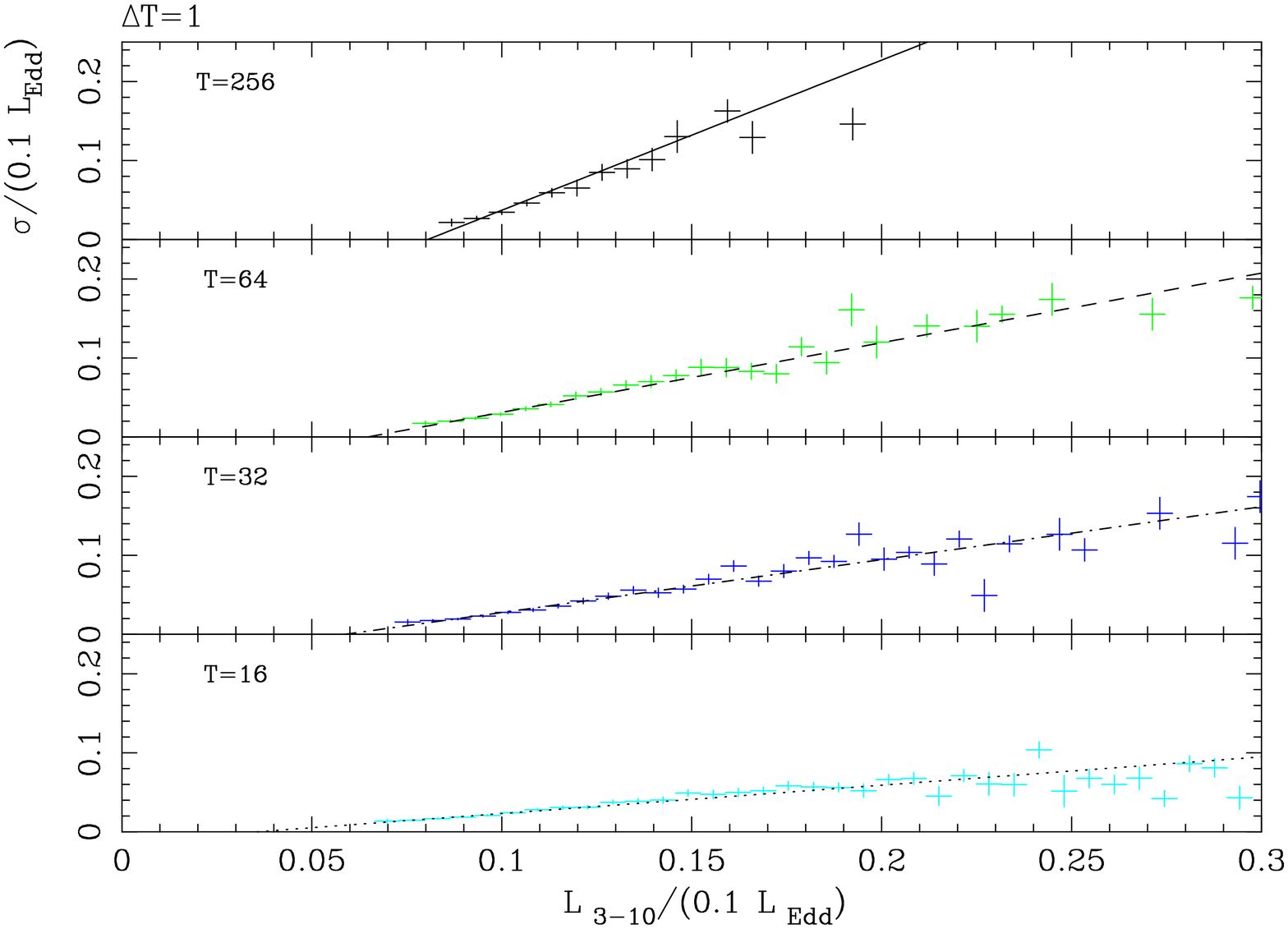, width=8cm}
  \end{center}
\caption{Luminosity dependence of the variance calculated for segments
of a simulated lightcurve ($D=1$, $C=10^{-3}$ and $\tau=1.5$) with a
fixed temporal resolution of $60$~seconds. 
The duration of the different segments (in minutes) is written in each 
panel. The fitting functions are of the form $\sigma =
k(L_{3-10}-L_0)$, with $k$ and $L_0$ constants.}
\label{amerloni-C2_fig_dt}
\end{figure}

Then, keeping fixed the time resolution, we calculated 
mean and variance
for segments of different length. The results, binned in luminosity intervals,
are shown in Fig.~\ref{amerloni-C2_fig_dt}, where the length of the segments $T$ is
measured in minutes. The variance  corresponds to the integral
of the PDS shown in Fig.~\ref{amerloni-C2_pds} from frequencies $1/(T \times 60)$ to $1/60$ (Hz).

A clear correlation is evident in all cases,
consistent with a linear one with an offset. 
\cite*{amerloni-C2:umh01} argues that the correlation
rules out any possible shot noise model. Still, 
we are able to reproduce it within our thundercloud model, which is
indeed a generalized shot noise model. 
One possible explanation for this lies in the way the lightcurve is `randomized'.
Once the average covering fraction $C$ and the distributions indexes
$D$ and $p$ are fixed,
the total expected number of active region at any time $N_{\rm tot}(t)$ is calculated, 
then the {\it actual} (randomized)  number, $\cal N$$_{\rm tot}(t)$,  
is picked from a poissonian distribution that has $N_{\rm tot}(t)$ as its
mean. Then, starting from the smallest flares, we calculate in similar fashion the number 
of active regions of every size in such a way as to ensure that their total is exactly
$\cal N$$_{\rm tot}(t)$. Therefore, random variations in $\cal N$$_{\rm tot}(t)$ 
(which correspond to random variations in the covering fraction $C$) 
are transferred to flares of all sizes, and of all timescales,
thus enforcing a correlation between the overall flux level and the amount of
variability on all time-scales. This merely reflects the fact that
large active regions are indeed made of a collection of correlated
microflares (avalanches). 

\section{Long timescales: low-luminosity systems}
\label{amerloni-C2_sec_long}

On timescales much longer than the dynamical one, the system evolution
is regulated by the global properties of the accretion flow. Here we
want to discuss the implication of our coronal model for the long-term 
properties of systems accreting at low rates (as compared to the
Eddington one). In particular we will consider Galactic Black Hole Candidates  (GBHC) in their {\it low/hard}
state and the so-called Low Luminosity AGN (LLAGN). These two classes
of systems have many common features:
GBHC have very weak (or absent) quasi-thermal 
spectral component in the soft X-ray/EUV spectral range; 
hard (photon index $\Gamma = 1.5 -2$) X-ray  power-law spectra, 
rolling over at $\sim 100$~keV 
(evidence of thermal Comptonization in a optically thin medium); 
compact (unresolved) flat or inverted radio core, probably extending
to the NIR or optical regime (\cite{amerloni-C2:fen01}) and a clear {\it temporal}
correlation between the radio and hard X-ray fluxes.
LLAGN, in turn, show a remarkable weakness of the 
Big Blue Bump (\cite{amerloni-C2:ho99}); hard X-ray spectra ($\Gamma = 1.6-1.8$); 
compact radio cores nearly ubiquitously, often accompanied by 
`jet-like' features; radio spectra with turnover frequencies 
of the order $\sim 0.1-100$~GHz and radio loudness anti-correlated 
with accretion rate 
(for nearby ellipticals with reliable mass estimates; \cite{amerloni-C2:ho02}).

All these observational evidences suggest that outflows are common 
features of low $\dot m$ systems with hard X-ray spectra. 

\subsection{Corona dominated accretion discs}
As shown in Merloni \& Fabian (2002, MF2), if angular momentum transport in a standard,
geometrically thin and optically thick accretion disc
is due to magnetic turbulent stresses, the magnetic energy density and
the effective viscous stresses inside the disc are proportional
to the geometric mean of the total (gas plus radiation) and gas
pressure:
\begin{equation}
\frac{B^2}{8\pi}=P_{\rm mag}\simeq \alpha_0 \sqrt{P_{\rm tot}P_{\rm
gas}},
\end{equation}
with $\alpha_0$ constant (of the order of unity).

Then, the fraction of gravitational power $f$ 
released in a magnetic corona increases as the accretion rate
decreases,
because the disc is more and more gas pressure dominated even in its
inner parts. When the disc is completely gas pressure dominated 
(at accretion rates smaller than the critical value $\dot m_{\rm crit}
\sim 0.1 (\alpha_0 M_{\rm BH}/M_{\sun})^{-1/8}$), the strength of the corona depends on
the poorly understood mechanisms of vertical flux tube transport in
the disc, but it is likely that $f>0.7$ (see MF2 for a
more thoroughly discussion).

\subsection{MHD jets from magnetic coronae}
The magnetic energy density in the corona is given by
\begin{equation}
\label{amerloni-C2_eq_b_lh}
\frac{B^2}{8\pi}=\frac{3 L_{\rm H}}{4\pi R_a^2 N_{\rm tot} c}\left(\frac{c}{v_{\rm diss}}\right),
\end{equation}
where $v_{\rm diss}$ is the dissipation velocity and depends on the uncertain nature of the
reconnection process (MF1), and can be assumed to be of the order
$v_{\rm diss} \sim 0.01 c$.

Models and simulations of jet production (\cite{amerloni-C2:bz77,amerloni-C2:me99}) show that it
is the {\it poloidal} component of the magnetic field which mainly drives the
production of powerful jets, and the output power
 can be expressed as (Livio, Ogilvie \&
Pringle 1999)
\begin{equation}
\label{amerloni-C2_eq_lj}
L_{\rm j}=\left(\frac{B_{\rm p}^2}{8\pi}\right)2\pi R_{\rm cor}^2 R_{\rm cor}\Omega.
\end{equation}
Here $R_{\rm cor}$ is the size of the region where most of the coronal power
is dissipated and $\Omega$ is the typical angular velocity of the magnetic
field lines.

If $H_a$ is the typical coronal flux tube scaleheight (height of a reconnection site), we have
$B_{\rm p}/B \simeq H_a/R_{\rm cor}$, and therefore
\begin{equation}
\label{amerloni-C2_eq_lj_lh}
L_{\rm j}=\frac{3}{2}L_{\rm H} \left(\frac{c}{v_{\rm diss}}\right)
\left(\frac{H_a}{R_{\rm cor}}\right)^2\left(\frac{R_{\rm cor} \Omega}{c}\right),
\end{equation}
which in turn gives, for the fraction of coronal power that goes into the MHD outflow
\begin{equation}
\label{amerloni-C2_eq_eta}
\eta=\left(1+\frac{2}{3}\left(\frac{v_{\rm diss}}{c}\right)
\left(\frac{R_{\rm cor}}{H_a}\right)^2\left(\frac{c}{R_{\rm cor} \Omega}\right)\right)^{-1}.
\end{equation}
By inspection of Eq.(\ref{amerloni-C2_eq_eta}) we can conclude that the outflow power is stronger if
the coronal scaleheight is large with respect to the distance from the central source. 
This would help in increasing
the relative strength of the poloidal component of the magnetic field, that is 
the one ultimately responsible for the powering of the jet. 

As an example, for $v_{\rm diss} \simeq 0.01 c$, $R_{\rm cor}\sim
7R_{\rm S}$, $H_a\sim 2R_{\rm S}$ and
$\Omega=\Omega_{\rm K}(R_{\rm cor})$, we obtain $\eta_{\rm MHD} \simeq 0.55$: the MHD jet can carry
away a substantial fraction of the coronal power.

\section{Conclusions}
We have presented a model to explain spectral and temporal variability
on the smallest timescales in the X-ray emission from Seyfert Galaxies
and GBHC. We have simulated X-ray light-curves that reproduce the
observed PDS properties and the spectral variability. The basic
geometric properties of the corona we propose are the following: 
\begin{itemize}
\item{The corona must not be uniform, but structured and heated
intermittently (flares);}
\item{The spatial and temporal distribution of the flares are not
random, but proceed in correlated trains of events (avalanches);}
\item{The size of the avalanches determines the size of the active
regions, which are distributed as a power-law; larger avalanches are
more luminous and have softer spectra.}
\end{itemize}

On longer timescales, the evolution of the corona is governed by the
evolution of the accretion rate and/or of the inner disc geometry. At
low accretion rate, the strength of a magnetic corona produced by
buoyant magnetic flux tube amplified in an underlying standard
accretion disc increases.
If the energy in the corona, as we suggest, is stored in the magnetic
field, and the height of a reconnection site is much 
larger than its size, which is of the order of the disc
thickness, powerful MHD outflows can be launched from the inner corona.
 
\begin{acknowledgements}
ACF thanks the Royal Society for support. AM acknowledges the financial
support of PPARC and of the TMR research network ``Accretion onto
black holes, compact stars and protostars'', funded by the European
Commission under contract number ERBFMRX-CT98-0195.
\end{acknowledgements}


\begin{thebibliography}{}

\bibitem[\protect\astroncite{Blandford \& Znajek }{1977}]{amerloni-C2:bz77}
Blandford, R.D., Znajek, R.L., 1977, MNRAS, 179, 433.

\bibitem[\protect\astroncite{Di Matteo, Celotti \& Fabian
}{1999}]{amerloni-C2:dcf99}
Di Matteo, T., Celotti, A., Fabian, A.C., 1999, MNRAS, 304, 809.

\bibitem[\protect\astroncite{Done }{2001}]{amerloni-C2:don01}
Done, C., 2001, Adv. Space Res., 28, 255.

\bibitem[\protect\astroncite{Fender }{2001}]amerloni-C2:{fen01}
Fender, R.P., 2001, MNRAS, 322, 31.

\bibitem[\protect\astroncite{Haardt, Maraschi \& Ghisellini
}{1994}]{amerloni-C2:hmg94}
Haardt, F., Maraschi, L., Ghisellini, G., 1994, ApJ, 432, L95.

\bibitem[\protect\astroncite{Ho }{1999}]{amerloni-C2:ho99}
Ho, L.C., 1999, ApJ, 516, 672.

\bibitem[\protect\astroncite{Ho}{2002}]{amerloni-C2:ho02}
Ho, L.C., 2002, to appear in `Issues on unifications of AGNs',
eds. R. Maiolino, A. Marconi, N. Nagar (San Francisco: ASP). astro-ph/0110439.

\bibitem[\protect\astroncite{Livio, Ogilvie \& Pringle}{1999}]{amerloni-C2:lop99}
Livio, M., Ogilvie, G.I., Pringle, J.E., 1999, ApJ, 512, 100.

\bibitem[\protect\astroncite{Meier}{1999}]{amerloni-C2:me99}
Meier, D.L., 1999, ApJ, 522, 753.

\bibitem[\protect\astroncite{Merloni}{2001}]{amerloni-C2:me02}
Merloni, A., 2001, PhD thesis.

\bibitem[\protect\astroncite{Merloni \& Fabian}{2001}]{amerloni-C2:mf01}
Merloni, A., Fabian, A.C., 2001b, MNRAS, 328, 958 (MF1).

\bibitem[\protect\astroncite{Merloni \& Fabian}{2002}]{amerloni-C2:mf02}
Merloni, A., Fabian, A.C., 2001a, MNRAS, in press (MF2).

\bibitem[\protect\astroncite{Nandra }{2001}]{amerloni-C2:nan01}
Nandra, K., 2001, astro-ph/0012448. 

\bibitem[\protect\astroncite{Poutanen \& Fabian}{1999}]{amerloni-C2:pf99}
Poutanen, J., Fabian, A.C., 1999, MNRAS, 306, L31.

\bibitem[\protect\astroncite{Stern et al.}{1995}]{amerloni-C2:ste95}
Stern, B.E., Poutanen, J., Svensson, R., Sikora, M., Begelman, M.C.,
1995, ApJ, L13. 

\bibitem[\protect\astroncite{Uttley \& McHardy}{2001}]{amerloni-C2:umh01}
Uttley, P., McHardy, I.M., 2001, MNRAS, 323, L26.

\bibitem[\protect\astroncite{Vaughan \& Edelson}{2001}]{amerloni-C2:ve01}
Vaughan, S.A., Edelson, R., 2001, ApJ, 548, 694.

\end{thebibliography}
\end{document}